\documentclass[12pt]{iopart}

\usepackage{graphicx}%
\usepackage{multirow}%
\usepackage{amssymb,amsfonts}%
\usepackage{amsthm}%
\usepackage{hyperref}
\usepackage{url}

\begin{document}

\title[Leggett-Garg inequality and neutrino oscillations]{Evaluation of the Leggett-Garg inequality by means of the neutrino oscillations observed in reactor and accelerator experiments}

\author{Ricardo Zamora Barrios\footnote{Present address: Departamento de Física, Universidad Catolica del Norte, Avenida Angamos 0610, Casilla 1280, Antofagasta, Chile}, Mario A.~Acero}

\address{Programa de Física, Universidad del Atlántico, Carrera 30 No.~8-49, Puerto Colombia, Atlántico, Colombia}

\begin{abstract}
We revisit the study of the violation of the Leggett-Garg inequality in neutrino oscillation data as a means to test some of the fundamental aspects of quantum mechanics. In particular, we consider the results of the Daya Bay and RENO reactor experiments, and the MINOS and NOvA accelerator experiments. We find that DB and MINOS exhibit a strong manifestation of Leggett-Garg violation, whereas for RENO and NOvA data, the indication is weaker. Considering the particular baselines and energy ranges explored by each experiment, our results demonstrate that the Leggett-Garg violation is more evident for smaller baseline-to-energy ratios in all the data sets studied, a relevant aspect to consider when looking for evidence of quantum mechanical decoherence in neutrino oscillations.
\end{abstract}

%
\vspace{2pc}
\\
%
\submitto{\JPG}
%

\section{Introduction}\label{sec_Intro}
Neutrino physics has an impact on a wide range of aspects of particle physics, astrophysics, and cosmology. In particular, the well established phenomenon of neutrino oscillations \cite{Super-Kamiokande:1998kpq,Super-Kamiokande:2002ujc,SNO:2002hgz,KamLAND:2002uet,T2K:2011ypd,DoubleChooz:2011ymz,DayaBay:2022orm,RENO:2018dro,MINOS:2020llm,NOvA:2021nfi}, demonstrating that neutrinos are massive particles, constitutes one of the first evidence of physics beyond the Standard Model of particle physics. In addition, this phenomenon, has also granted the opportunity to study the violation of the Charge-Parity CP symmetry in the lepton sector, which is thought to account for the matter-antimatter asymmetry of the Universe \cite{Pilaftsis:1997jf,Buchmuller:2004nz,Buchmuller:2005eh}.

On the other hand, being a quantum mechanical process, it has been recently noticed \cite{Formaggio:2016cuh} that neutrino oscillations are also a useful tool to look at some of the fundamental aspects of Quantum Mechanics (QM) at a macroscopic level, thanks to the large-scale nature of the phenomenon, i.e., the neutrino flavor transformation takes place while neutrinos travel long distances (from $\mathcal{O}(10^2)$ m to $\mathcal{O}(10^6)$ km). 

The relevance and impact of QM in science and technology have motivated scientists to search for ways to test its foundations. One of those tests is the well-known Bell's Inequality (BI) \cite{Bell:1964kc}, which provided a mathematical formulation to study local realism, i.e., a property of \emph{classical} systems, according to which any observable would have a definite value at any moment, and that independent results are obtained when individual measurements of the observables are performed at different, \emph{spatially} separated locations. Several experimental setups have been implemented \cite{Aspect:1981nv,Weihs:1998gy,Hensen:2015ccp,Giustina:2015yza} and have observed a violation of the BI, supporting the idea of nonlocality, thus demonstrating the quantum nature of the systems.

In analogy to the BI, Leggett and Garg \cite{Leggett:1985zz} proposed an approach allowing to implement the test by performing measurements of the observables at different \emph{times}, opening the possibility to apply the test to macroscopic systems. As pointed in Ref.~\cite{Formaggio:2016cuh}, with this so-called Leggett-Garg Inequality (LGI), we are able to prove the foundations of QM, without going into some of the difficulties observed when applying tests of the BI.

As highlighted in Ref.~\cite{Emary:2014rep}, Leggett and Garg were also looking for the possibility of studying macroscopic coherence in the laboratory, based on two principles which describe the macroscopic word: \emph{i)} macroscopic realism (MR), according to which making a measurement on a macroscopic system results in a well-defined preexisting value, and \emph{ii)} non-invasive measurability (NIM), indicating that the performed measurement does not disturb the system. The key point here is that classical mechanics complies with these premises, but QM does not: \emph{i)} would be violated by the existence of a macroscopic superposition, and its quantum-mechanical collapse under measurement violates \emph{ii)} \cite{Emary:2014rep} (for additional references and details on this topic, see also Refs.~\cite{Naikoo:2019eec,Wang:2022tnr}). With this in mind, the LGI can be used to study some of the foundations of QM at a macroscopic level and to test the existence of macroscopic coherence: the violation of the LGI would indicate that \emph{i)} or \emph{ii)} (or both) is untrue and a macroscopic description of the system would not be admissible.

In this paper, after briefly describing the phenomenon of neutrino oscillations and commenting about the oscillation probability as obtained from QM in Sec.~\ref{sec_NoOsc}, we present an updated study of the violation of the LGI in the neutrino oscillation phenomenon, by looking at the experimental data from the electron antineutrino disappearance collected by Daya Bay (DB) \cite{DayaBay:2022orm} and RENO \cite{RENO:2018dro} reactor neutrino experiments, and from the muon neutrino disappearance compiled by the MINOS \cite{MINOS:2020llm,Evans:2017brt,Sousa:2015bxa} and NOvA \cite{NOvA:2021nfi,Catano-Mur:2022kyq} long baseline (LBL) accelerator neutrino experiments (described in Sec.~\ref{sec_Experiments}).

The rest of the article covers the necessary information and descriptions to advance the investigation and includes an explanation of the LGI and the parameters used in this study in Sec.~\ref{sec_LGI}. The final results are then explained and discussed in Sec.~\ref{sec_Results}, and the conclusions are drawn in Sec.~\ref{sec_Conclusion}.

\section{Neutrino oscillations}\label{sec_NoOsc}
The mathematical description of neutrino oscillation from the perspective of quantum mechanics is based on the linear superposition of flavor eigenstates, $| \nu_{\alpha} \rangle$ ($\alpha = e,\mu,\tau$), in terms of mass eigenstates, $| \nu_{i} \rangle$ ($i = 1,2,3$), through the Pontecorvo–Maki–Nakagawa–Sakata (PMNS) matrix $U$ \cite{Maki:1962mu,Pontecorvo:1967fh}, as 
\begin{equation}
| \nu_{\alpha} \rangle = \sum_{i} U^{*}_{\alpha i}|\nu_{i} \rangle.
\label{eq:Superposition}
\end{equation}
The time evolution of flavor neutrinos is then governed by that of massive neutrinos, so that \cite{Giunti:2007ry}
\begin{equation}
|\nu_{\alpha}(t)\rangle = \sum_{i}^{}U^{*}_{\alpha i}  e^{-iE_{i}t} |\nu_{i}\rangle,
\label{eq:ChangeOfFlavorInTime2}
\end{equation}
where $E_i$ is the energy associated with the eigenstate $| \nu_i\rangle$. Since the PMNS matrix is unitary, from Eq.~(\ref{eq:Superposition}) one can express the mass states as a superposition of flavor states, leading to
\begin{equation}
|\nu_{\alpha}(t)\rangle = \sum_{\beta=e,\mu,\tau} \left( \sum_{i}U^{*}_{\alpha i}e^{-iE_{i}t} U_{\beta i} \right) |\nu_{\beta} \rangle.
\label{eq:trans}
\end{equation} 
From here it is possible to compute the probability that a neutrino created in a particular flavor state $\alpha$, changes to a different flavor $\beta$, after some time $t$, given by $P_{\nu_{\alpha} \rightarrow \nu_{\beta}}(t) = \left| \left\langle \nu_\beta | \nu_\alpha \right\rangle \right|^{2}$. Considering the ultrarelativistic nature of neutrinos, and a coherent evolution of the neutrino mass eigenstates (decoherence effects at the considered baselines can be neglected), the oscillation probability is 
\begin{equation}
P_{\nu_{\alpha} \rightarrow \nu_{\beta}}(L,E) = \sum_{i,j} U_{\alpha i} U_{\beta i}^{*} U_{\alpha j}^{*} U_{\beta j} e^{-i\left(\frac{\Delta m_{ji}^{2}}{2E}L\right)},
\label{eq:pro2}
\end{equation}
where $\Delta m_{ji}^{2}$ represents the squared-mass difference, $\Delta m_{ji}^{2} \equiv m_{j}^2 - m_{i}^2$, and we have set $t=L$ (with $c=1$), whit $L$ being the distance traveled by the neutrino from its source to the detector locations.

Neutrino masses, and then $\Delta m_{ji}^{2}$, together with the PMNS matrix elements $U_{\alpha i}$, are physical constants given by Nature and constitute the oscillation parameters to be measured through experimentation; however, the distance traveled by the neutrinos $L$, and their energy $E$, are defined by the experiment design. If we define oscillation phase
\begin{equation}
    \Phi_{ji} = - \left(\frac{\Delta m_{ji}^{2}\,L}{2E}\right),
    \label{eq:OscPhase}
\end{equation}
the particular configuration of each experiment, setting specific values for $L$ and $E$, allows probing different oscillation regions, as it will become more clear later.

\subsection{The case of two flavor neutrinos}
A useful and precise approximation is to consider a  simplified framework where two neutrino flavors are considered. Under this effective model, the oscillation probability simplifies considerably and, it turns out, has been demonstrated to be a good estimation to analyze most of the experimental data, given that no experiment is truly sensitive to all the parameters governing the more realistic three-neutrino mixing phenomenon.

Specifically, in this model, the mixing matrix is a $2\times 2$ orthogonal matrix, depending on one single mixing angle $\theta$. Also, there is a unique squared-mass difference, $\Delta m^2$, so the oscillation probability, as given by Eq.~(\ref{eq:pro2}), reduces to
\begin{equation}
P_{\nu_\alpha \rightarrow \nu_\beta}(L, E) = \sin^22\theta \, \sin^2\left(\frac{\Delta m^2 L}{4 E}\right) \qquad (\alpha \neq \beta),
\label{eq:prob2neutrinos2}
\end{equation}
from which the survival probability, i.e., the probability that, after traveling a certain distance $L$, the final flavor neutrino state is the same as the initial one, becomes
\begin{equation}
    P_{\nu_{\alpha} \rightarrow \nu_{\alpha}}(L, E) = 1 - \sin ^2 2 \theta \sin ^2\left(\frac{\Delta m^2 L}{4 E}\right).
    \label{eq:survivalprobability2}
\end{equation}
Expressions (\ref{eq:prob2neutrinos2}) and (\ref{eq:survivalprobability2}) are the basis of the analysis presented here as will become clear later, considering that in the context of the selected experimental data, the violation of the LGI is studied for the so-called appearance (oscillation probability) or disappearance (survival probability) channels. 

\section{Leggett-Garg Inequality}\label{sec_LGI}
The two-neutrino oscillation framework explained in the previous section constitutes a suitable dichotomous (i.e.,  two level) system to test the LGI. Indeed, as the initial neutrino state $\nu_{\mu}$ ($\nu_{e}$) evolves in space-time, it only has the possibility to transform into a different flavor state, $\nu_{e}$ ($\nu_{\mu}$) or to remain unchanged.

It is in this sense that we say that the system can only take two values: $Q = +1$ (flavor does not change) or $Q = -1$ (flavor does change). These values are measured in a non invasive manner at different times, $t_{i}$, and the correlation between the measurements at times $t_{i}$ and $t_{j}$ is computed as \cite{Formaggio:2016cuh}
\begin{equation}
    C_{i j} \equiv\left\langle\widehat{Q}\left(t_{i}\right) \widehat{Q}\left(t_{j}\right)\right\rangle,
    \label{eq:correlation}
\end{equation}
where $\langle\,\cdots\rangle$ represents the average over the performed tests which can be calculated by
\begin{equation}
    C_{i j}=\sum_{\widehat{Q}_{i} \widehat{Q}_{j}=\pm 1} \widehat{Q}_{i} \widehat{Q}_{j} \, \mathbb{P}_{\widehat{Q}_{i} \widehat{Q}_{j}}\left(t_{i}, t_{j}\right),
    \label{eq:correlation2}
\end{equation}
with $\mathbb{P}_{\widehat{Q}_{i} \widehat{Q}_{j}}\left(t_{i}, t_{j}\right)$ being the joint probability of obtaining the results $\widehat{Q}_{i}$ and $\widehat{Q}_{j}$ from successive measurements at times $t_{i}$ and $t_{j}$, respectively \cite{Shafaq:2021lju}. Then, the LGI parameter, $K_{n}$, for measurements taken at $n$ different times, is defined as follows:
\begin{equation}
    K_{n} \equiv \sum_{i=1}^{n-1} C_{i, i+1}-C_{n, 1}.
    \label{eq:LGIparameter}
\end{equation}
Considering $n \geq 3$, realistic systems obey the LGI 
\begin{equation}
    K_{n} \leq n - 2.
    \label{eq:constantk2}
\end{equation}

In the particular cases where $n=3$  ($t_1 < t_2 < t_3$) and $n=4$ ($t_1 < t_2 < t_3 < t_4$), the corresponding parameters, $K_n$, given by equation (\ref{eq:LGIparameter}), are
\begin{equation}
    K_{3} = C_{1 2} + C_{2 3} - C_{1 3}, \qquad
    K_{4} = C_{1 2} + C_{2 3} + C_{3 4} - C_{1 4},
    \label{eq:k3_4}
\end{equation}
satisfying the inequalities
\begin{equation}
     K_{3} \leq 1,\qquad
     K_{4} \leq 2,
    \label{eq_ik34_}
\end{equation}
respectively. Therefore, a system is said to violate LGI if $K_3 > 1$ or $K_4 > 2$, indicating an infringement of one or both of the principles described in Sec. \ref{sec_Intro} (i.e., MR and NIM).

Using the correlations given by Eq.~(\ref{eq:correlation2}), we obtain $K_n$ in (\ref{eq:LGIparameter}) in terms of the neutrino survival probability of a neutrino of flavor $\alpha$, $P_{\alpha\alpha}$\footnote{There is a classical version for $K_n$, obtained assuming that $\widehat{Q}\left(t_{i}\right)$ and $\widehat{Q}\left(t_{j}\right)$ commute \cite{Formaggio:2016cuh}, which can be written as a function of the neutrino survival probability as
$$
  K_{n}^{c}  =2 \sum_{a=1}^{n-1}  P_{\alpha\alpha}(\psi_{a}) - \prod_{a=1}^{n-1} \left(2 P_{\alpha \alpha}(\psi_{a})-1\right)-(n-1).
$$}:
\begin{equation}
    K_n=(2-n)+2 \sum_{a=1}^{n-1} P_{\alpha \alpha}\left(\psi_{a}\right)-2P_{\alpha \alpha}\left(\sum_{a=1}^{n-1}\psi_{a}\right),
    \label{eq:Kn}
\end{equation}
where we have defined the phase $\psi_a=1.267\,\Delta m^2 \left(\frac{\Delta L}{E}\right)_a$.

For $n=3$ and $n=4$, i.e., for measurements performed at three and four different times, respectively, we consider a beam neutrinos composed by a particular flavor $\alpha \, (= e, \mu)$, in such a way that 
\begin{equation}
K_3 = -1 + 2\left[P_{\alpha\alpha}\left(\psi_1\right) + P_{\alpha\alpha}\left(\psi_2\right)\right] - 2 P_{\alpha\alpha}\left(\psi_1+\psi_2\right),
    \label{eq:k3_python}
\end{equation}
and
\begin{equation}
K_4 = -2 + 2\left[P_{\alpha\alpha}\left(\psi_1\right) + P_{\alpha\alpha}\left(\psi_2\right) + P_{\alpha\alpha}\left(\psi_3\right)\right] - 2P_{\alpha\alpha}\left(\psi_1+\psi_2+\psi_3\right).
    \label{eq:k4_python}
\end{equation}

To evaluate $K_3$ and $K_4$, each oscillation probability in Eqs.~(\ref{eq:k3_python}) and (\ref{eq:k4_python}) is selected (from each specific data set, as described later) so that the values of the phases $\psi_i$ follow the corresponding phase sum rule: $\psi_1 + \psi_2 = \psi_3$, for $K_3$ and $\psi_1 + \psi_2 + \psi_3 = \psi_4$, for $K_4$. However, as these sum rules might become too rigid considering the experimental observations, we apply a more general strategy. Specifically, the phases ($\psi_3$ or $\psi_4$) are obtained as follows: for the case of $K_3$, the sum $\psi_1 + \psi_2$ is calculated and then $\psi_3$ is determined by identifying the interval in which this sum falls, as shown in Figs.~(\ref{fig_Prob_React}) and (\ref{fig_Prob_Accel}). Subsequently, the corresponding probability associated with that bin is assigned. A similar procedure is applied for the case of $K_4$, based on the sum rule $\psi_1 + \psi_2 + \psi_3 = \psi_4$. In this updated approach, we do not impose strict constraints that the phases lie within a specific range around a reference phase. Instead, the phase correlations are intrinsically encoded in the binning procedure and the associated probabilities. 

\section{Selected experimental data}\label{sec_Experiments}
The test for LGI violations presented here is performed considering the data from the DB and RENO reactor neutrino experiments, and the MINOS and NOvA LBL accelerator neutrino experiments. 

\begin{figure}
 \centering
 \includegraphics[width=0.49\linewidth]{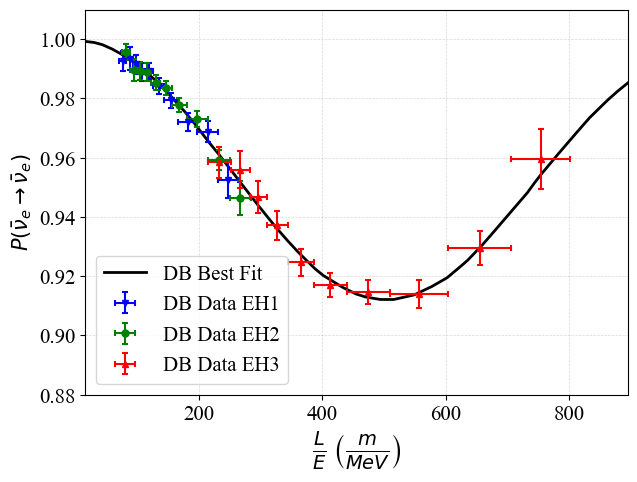}
 \includegraphics[width=0.49\linewidth]{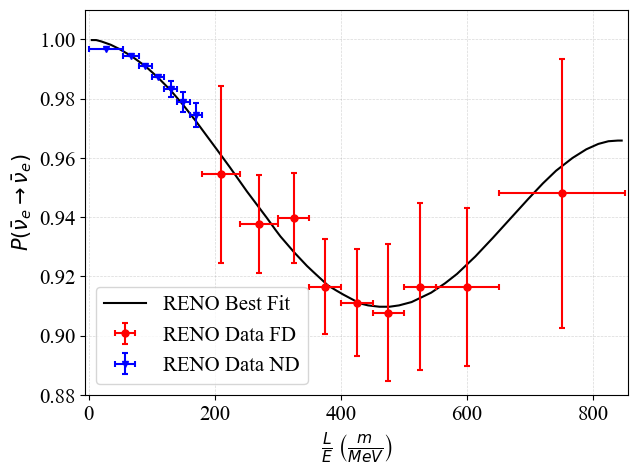}
 \caption{Experimental data and best fit prediction of the electron antineutrino survival probability as measured by DB \cite{DayaBay:2022orm} (\emph{left}) and RENO \cite{RENO:2018dro} (\emph{right}) reactor neutrino experiments.}
 \label{fig_Prob_React}
\end{figure}

Daya Bay and RENO collected data of the disappearance of electron antineutrinos, $\bar{\nu}_e$, produced by nuclear reactor cores. Daya Bay data includes the rate spectra from eight detectors located at three distances (identified as Experimental Halls, EH) from a set of six reactor cores, comprising baselines $L \in [360\,\rm{m}, 1985\,\rm{m}]$ (details on the detectors system and distances from the reactors can be found in Refs.~\cite{DayaBay:2015kir,Mezzetto:2010zi,Acero:2019fhe}), and with neutrino energies up to $\mathcal{O}(10)$ MeV \cite{DayaBay:2022orm}. In turn, RENO measured the $\bar{\nu}_e$ rate at two detectors located at a weighted baseline of 410.6 m (Near Detector, ND) and 1445.7 m (Far Detector, FD), respectively, with an energy up to 8 MeV \cite{RENO:2018dro} (see also Ref.~\cite{RENO:2016ujo} for additional details). Together with the prediction for the corresponding best fit (BF) oscillation parameters (black line), Fig.~\ref{fig_Prob_React} shows the data collected by the two reactor experiments.

\begin{figure}
 \centering
 \includegraphics[width=0.49\linewidth]{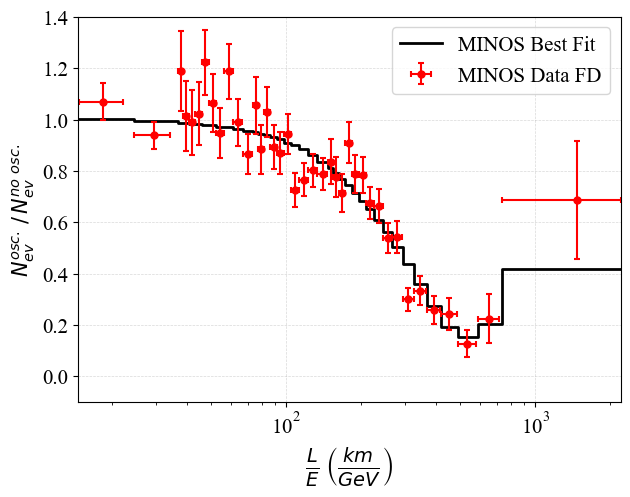}
 \includegraphics[width=0.49\linewidth]{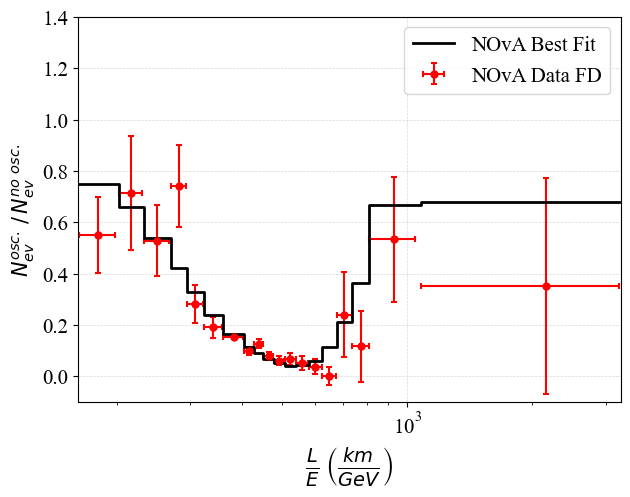}
 \caption{Experimental data and best fit prediction of the muon neutrino observed over expected (without oscillations) number of events, as measured by MINOS \cite{Sousa:2015bxa} (\emph{left}) and NOvA \cite{Catano-Mur:2022kyq} (\emph{right}) LBL accelerator experiments.}
 \label{fig_Prob_Accel}
\end{figure}

On the other hand, MINOS data \cite{MINOS:2020llm,Evans:2017brt,Sousa:2015bxa} were gathered by a detector located at a baseline $L = 735$ km from the $\nu_{\mu}$ source, with energy $E \sim 0.5-50$ GeV. Finally, NOvA observed the disappearance of $\nu_{\mu}$ after their travel to the detector at $L = 810$ km from the source with energy $E \sim 0.5-5$ GeV; we use the NOvA data as reported in Refs.~\cite{NOvA:2021nfi,Catano-Mur:2022kyq}. In a similar fashion to the reactor experiments, the prediction for the best fit oscillation parameters (black spectra) and the data collected by these two accelerator experiments (red dots with the corresponding error bars), are shown in Fig~\ref{fig_Prob_Accel}.

The plots in Figs.~\ref{fig_Prob_React} and \ref{fig_Prob_Accel}, clearly show a deviation from unity, meaning that all the experiments observed less neutrino events than expected. This variation is an indication of the disappearance of $\bar{\nu}_e$ (in DB and RENO) and $\nu_{\mu}$ (in MINOS and NOvA), respectively. Interpreted as the result of flavor oscillations, this is used here to test some of the fundamental aspects of QM by searching for evidence of LGI violations, as described in what follows. 

\section{Analysis and Results}\label{sec_Results}
In the first part of the analysis, Sec.~\ref{subsec_data}, we test the LGI using the experimental observations and the BF shown in Figs.~\ref{fig_Prob_React} and \ref{fig_Prob_Accel}. Then, in Sec.~\ref{subsec_pseudodata} we generate a number of pseudodata, considering the experimental uncertainties reported by each collaboration.
\subsection{Data and BF predictions}\label{subsec_data}
The first test is performed considering only the experimental data and the BF prediction of each experiment. In each case, the phase $\psi_i$ is built by taking $L/E$ at the center of the $i$-th bin 
Then, the associated oscillation probability is used to calculate $K_n$ with Eqs.~\ref{eq:k3_python} and \ref{eq:k4_python}.

\begin{figure}
 \centering
 \includegraphics[width=0.9\linewidth]{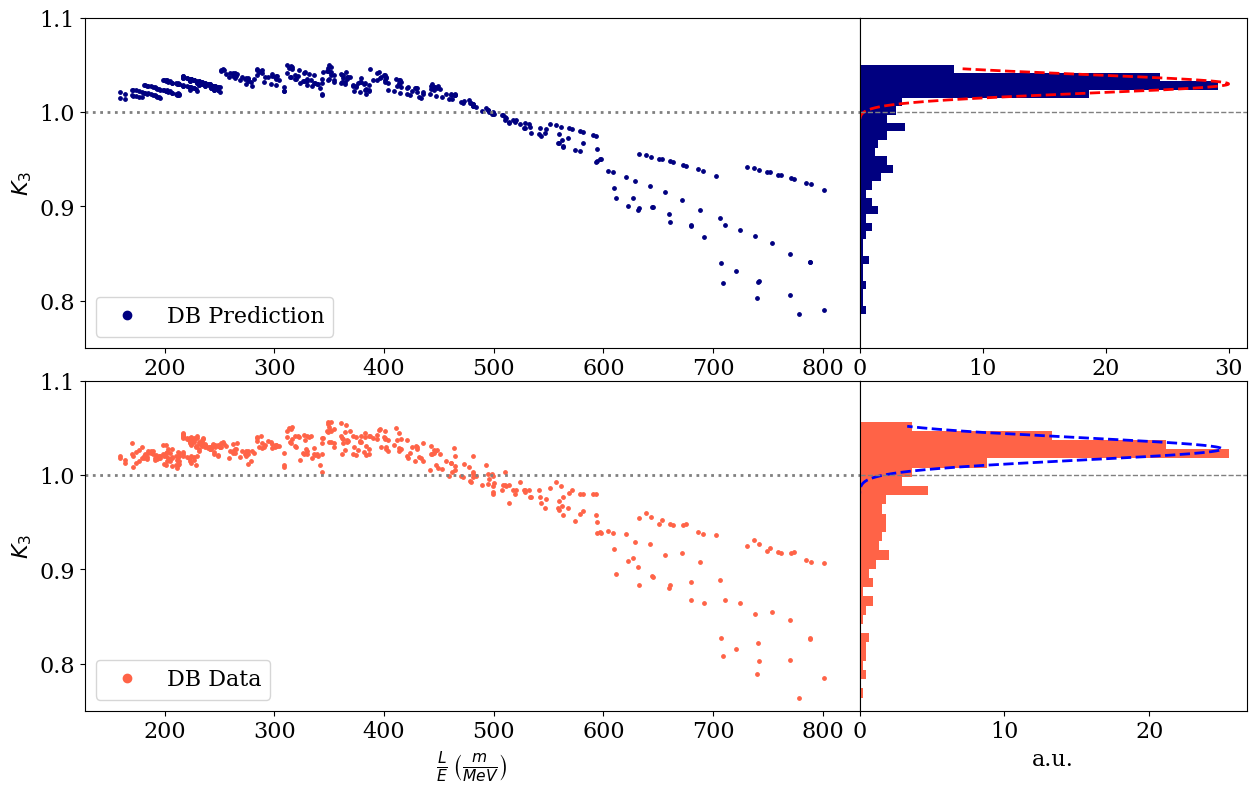}
 \includegraphics[width=0.9\linewidth]{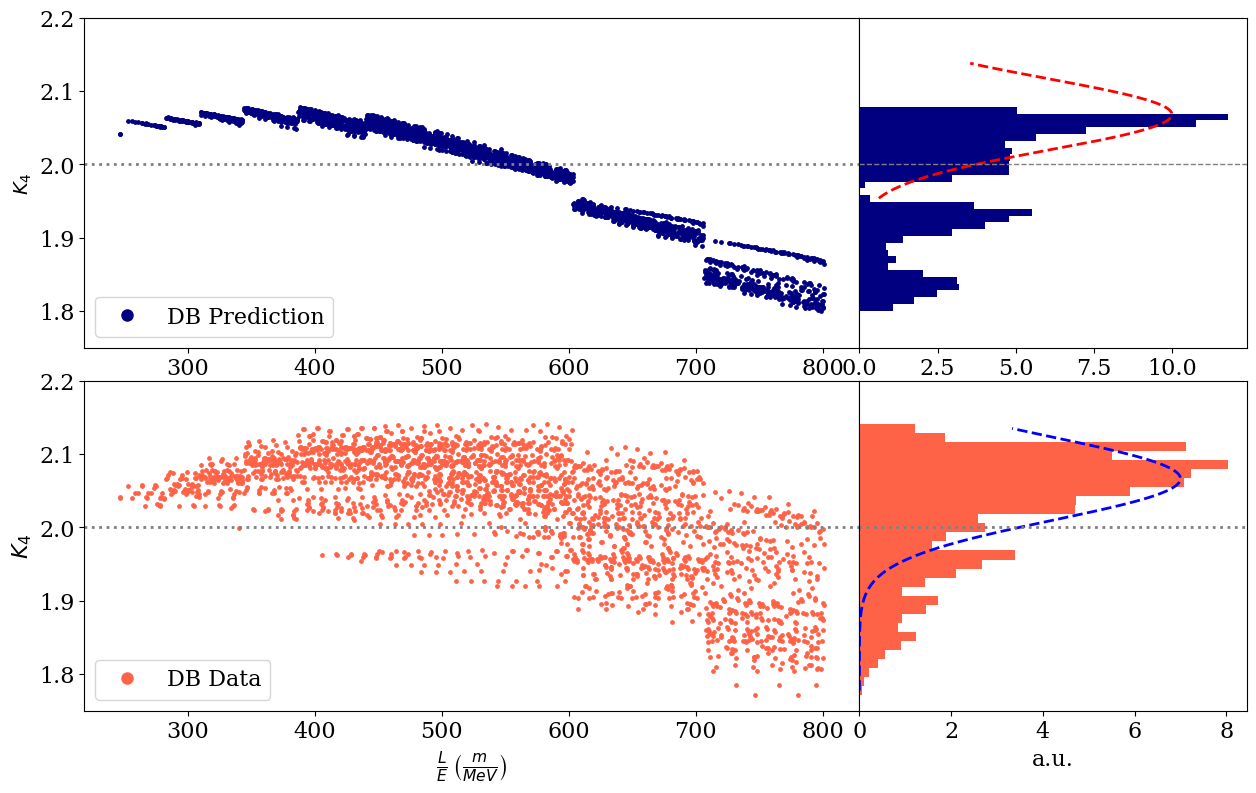}
 \caption{Points satisfying the LGI condition for $K_3$ (top) and $K_4$ (bottom) obtained from the Daya Bay BF prediction (blue dots and histograms) and the experimental data (red dots and histograms). The corresponding $K_{3,4}$ distributions are plotted as histograms on the right hand side, together with the fitted Gaussian.}
 \label{fig_k3DB}
\end{figure}

\begin{figure}
 \centering
 \includegraphics[width=0.9\linewidth]{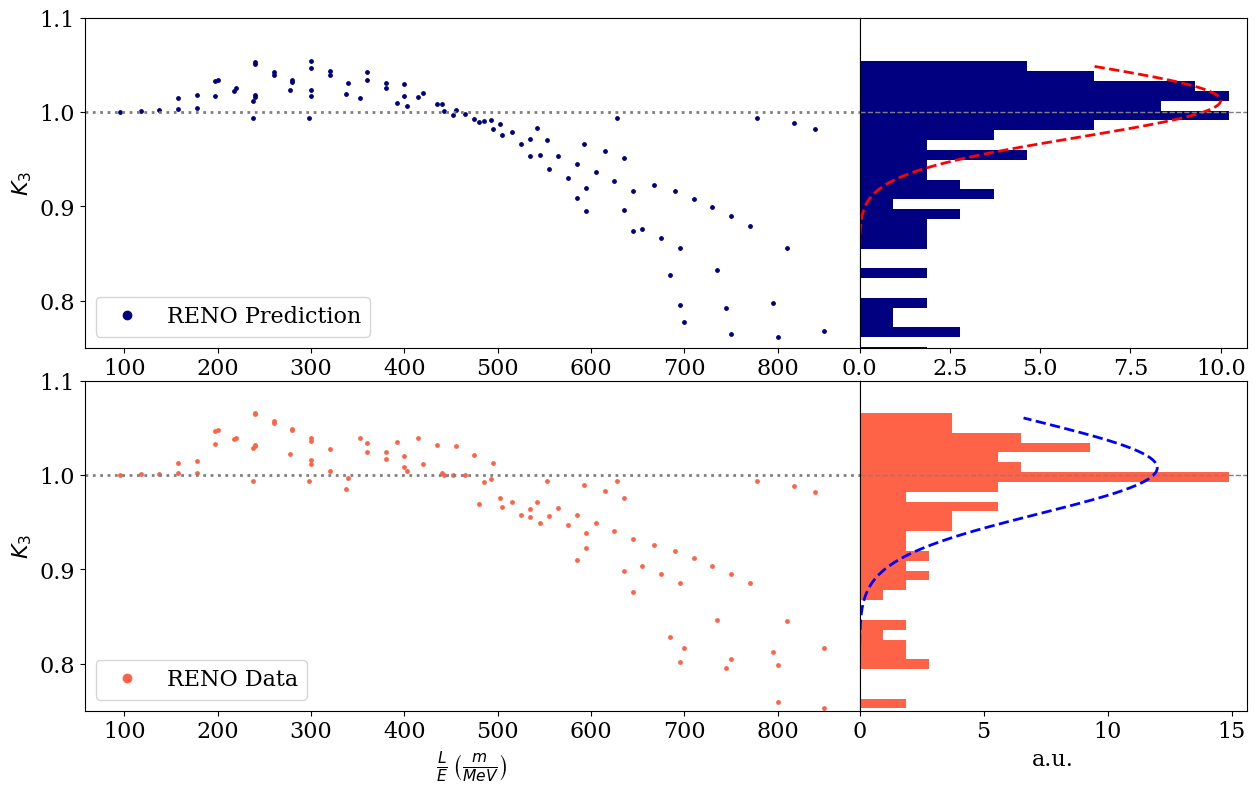}
 \includegraphics[width=0.9\linewidth]{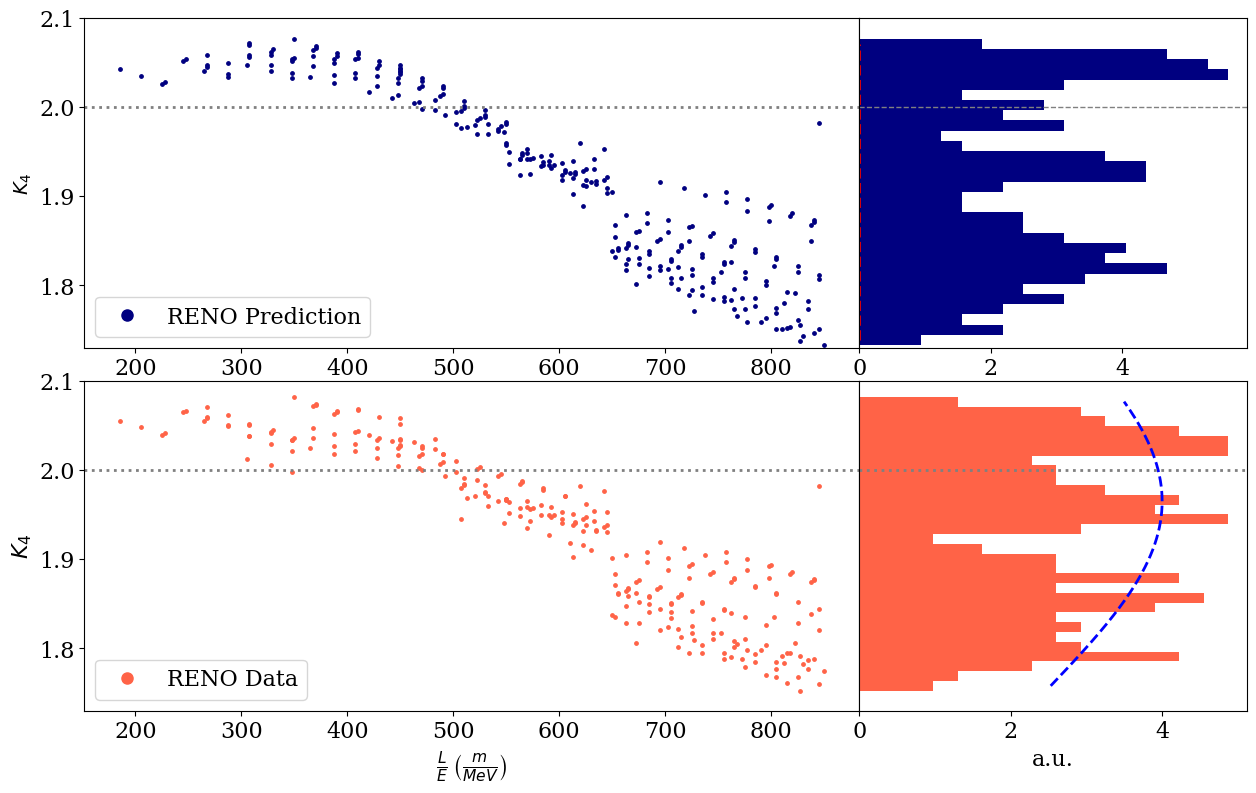}
 \caption{The same as in Fig.~\ref{fig_k3DB}, but for the RENO experiment.}
 \label{fig_k3RENO}
\end{figure}

\begin{figure}
 \centering
 \includegraphics[width=0.9\linewidth]{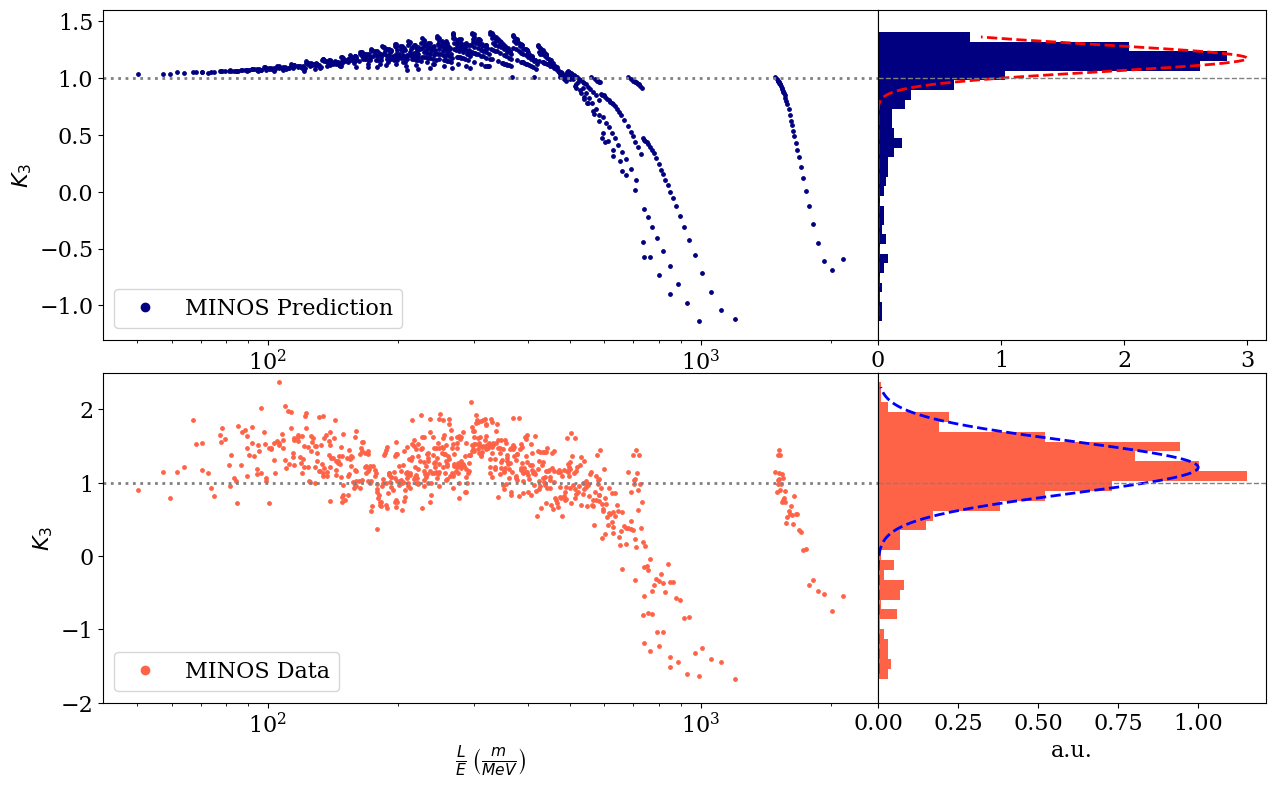}
 \includegraphics[width=0.9\linewidth]{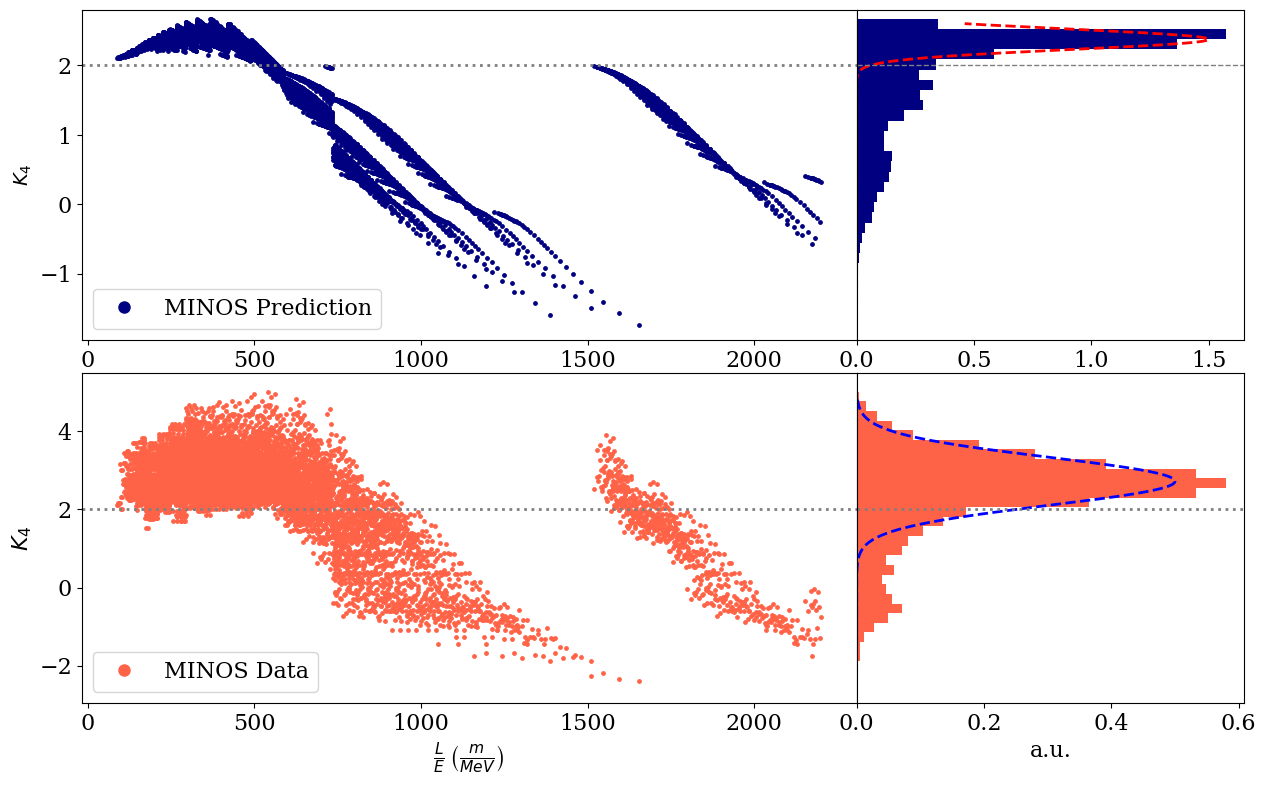}
 \caption{As in Fig.~\ref{fig_k3DB}, but for the MINOS experiment.}
 \label{fig_k3MINOS}
\end{figure}

\begin{figure}
 \centering
 \includegraphics[width=0.9\linewidth]{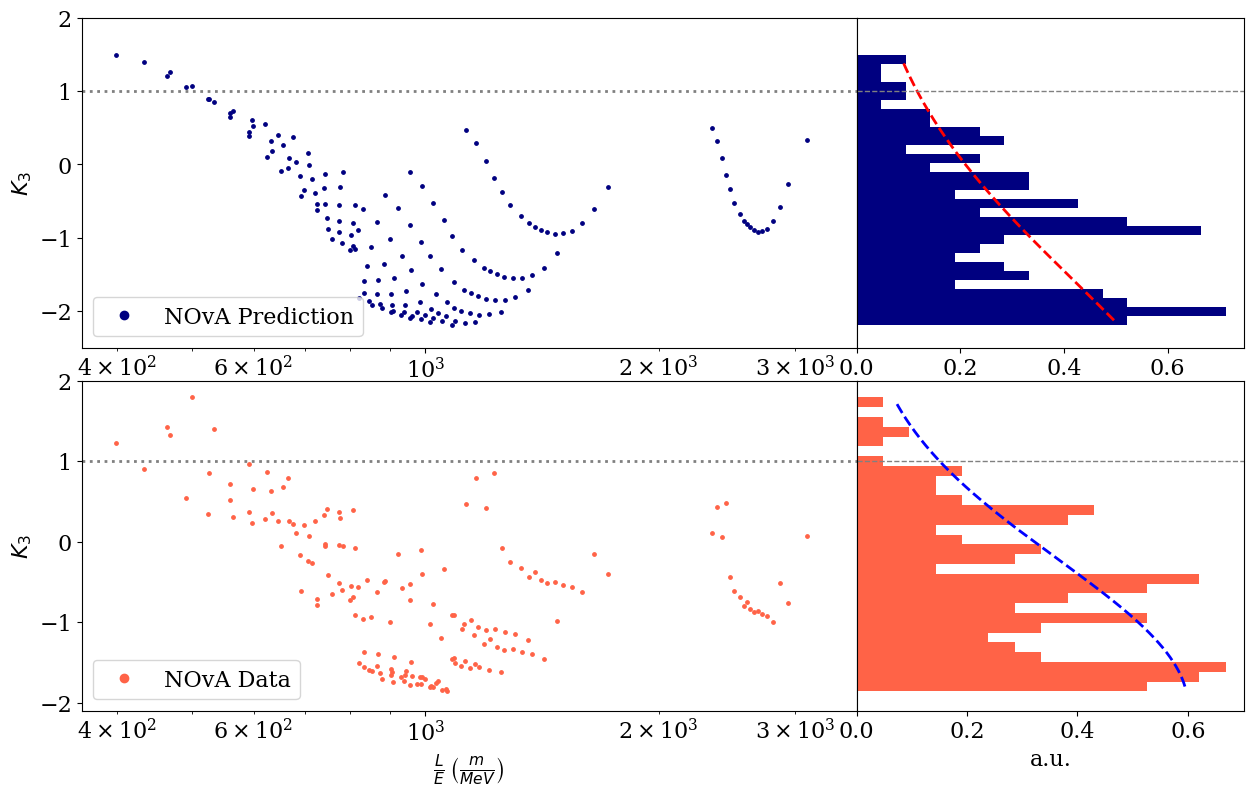}
 \includegraphics[width=0.9\linewidth]{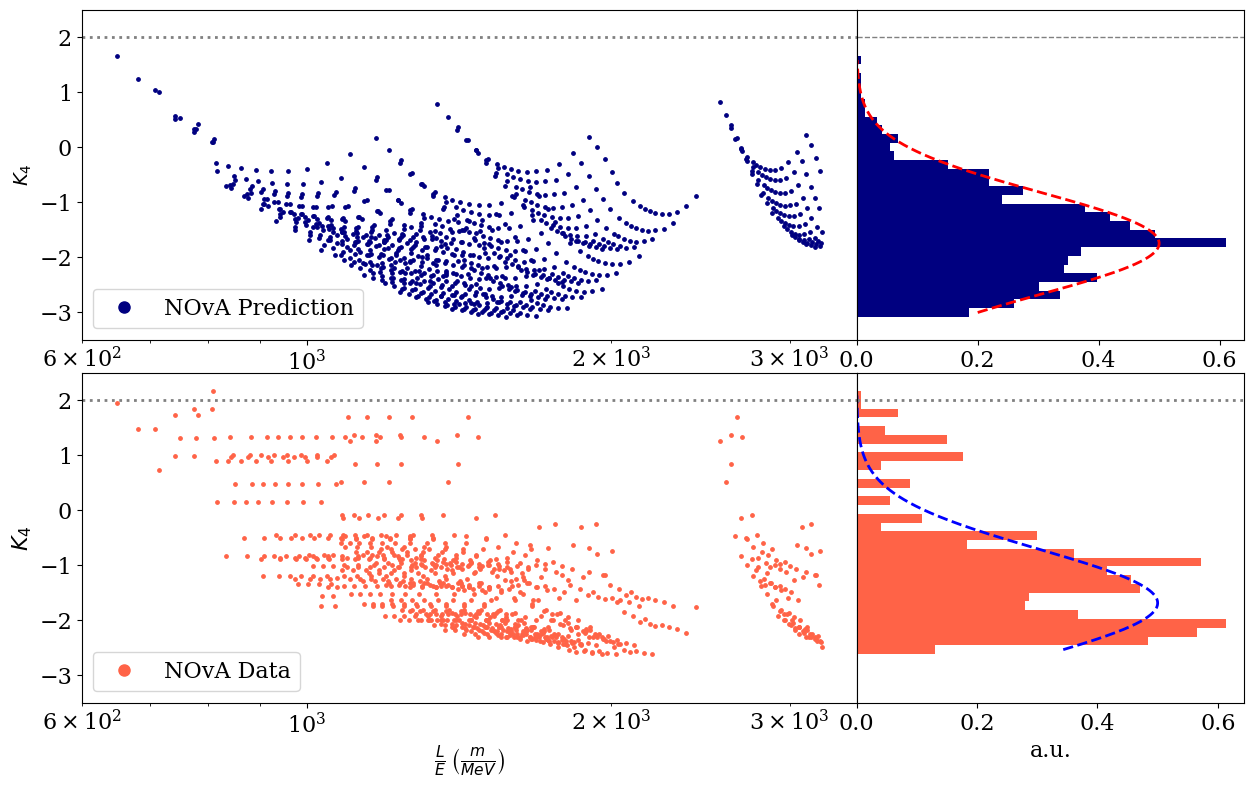}
 \caption{As in Fig.~\ref{fig_k3DB}, but for the NOvA experiment.}
 \label{fig_k3NOvA}
\end{figure}

The number of LGI violations ($K_3 > 1$ or $K_4 > 2$) is reported in Table \ref{tab:LGIViolations} for each experiment. Trying to better understand the general behavior of the LGI parameter, $K_n$, we attempt to fit a Gaussian probability distribution to the obtained values of the LGI parameters. The mean and standard deviation resulting from this fit are also shown in Table \ref{tab:LGIViolations}. In particular, columns 3 and 6 show the number of LGI violations compared with the total number of phases that satisfy the corresponding sum rule. As can be seen in the left panels of Figs.~\ref{fig_k3DB}--\ref{fig_k3NOvA}, some of the distributions of the number of LGI violations are not normal but exhibit a multi-peak structure, which reflects on poor or absent reported fit.
\begin{table}
\caption{Number of LGI violations obtained from analysis of each experiment. The mean and width of the fitted Gaussian distribution are also shown. The hyphen 
 (`-') indicates that it was not possible to fit a Gaussian to the obtained distribution.}\label{tab:LGIViolations}
\begin{tabular}{@{}llcccccc@{}}
\br
Experiment & & $K_3 > 1 \,(\%)$ & $\mu_3$ & $\sigma_3$ & $K_4 > 2 \,(\%)$ & $\mu_4$ & $\sigma_4$ \\
\mr
\multirow{2}{*}{DB}    & BF Pred. & 75.2 & 1.03  & 0.01 & 53.3 & 2.07  & 0.05 \\
                       & Data     & 73.2 & 1.03  & 0.01 & 70.9 & 2.07  & 0.06 \\\mr
\multirow{2}{*}{RENO}  & BF Pred. & 43.7 & 1.01  & 0.04 & 27.1 & -     & - \\
                       & Data     & 41.7 & 1.01  & 0.05 & 27.5 & 1.96  & 0.22 \\\mr
\multirow{2}{*}{MINOS} & BF Pred. & 77.3 & 1.17  & 0.12 & 59.6 & 2.37  & 0.15 \\
                       & Data     & 67.1 & 1.21  & 0.35 & 76.8 & 2.72  & 0.61 \\\mr
\multirow{2}{*}{NOvA}  & BF Pred. & 3.5  & -4.56 & 2.94 & 0.0  & -1.75 & 0.93 \\
                       & Data     & 2.9  & -2.05 & 1.83 & 0.1  & -1.69 & 0.98 \\
\br
\end{tabular}
\end{table}

The values of $K_3$ and $K_4$ in each phase $\psi_i \sim \left(\frac{L}{E}\right)_i$, are shown in Figs.~\ref{fig_k3DB} for BD, and \ref{fig_k3RENO} for RENO, Figs.~\ref{fig_k3MINOS} for MINOS and Fig.~\ref{fig_k3NOvA} for NOvA. The series of points for the same phase are obtained by considering the measurement error bars in Figs. \ref{fig_Prob_React} and \ref{fig_Prob_Accel}, which explains its scattered distribution. The horizontal dotted lines point to the corresponding LGI violation limit, $K_3>1$ and $K:4>2$, respectively.


The numbers in Table \ref{tab:LGIViolations} and the corresponding figures make it clear that the experimental results and the BF predictions of DB and MINOS show a strong indication of LGI violation; in fact, in the former case the violation occurs more than 70\,\% of the times for both $K_3$ and $K_4$, and the normal distributions with means $\mu_3 > 1$ (for $K_3$) and $\mu_4 > 2$ (for $K_4$), provide additional support to the above statement.

For RENO, the number of LGI violations is bellow $50\,\%$, and it is only possible to fit a normal distribution to the $K_3$ histogram (top panels of Fig.~\ref{fig_k3RENO}), with a mean just above 1 and a large enough width $\sigma_3$, allowing for a large number of points below the LGI violation condition. The $K_4$ distributions are too wide for a proper Gaussian fit, and most of the points obtained lie below $K_4 = 2$.

The case of NOvA is interesting because, as is clear from Fig.~\ref{fig_k3NOvA},
as well as Table \ref{tab:LGIViolations}, the number of LGI violation is $\lesssim 5\,\%$ (for $K_3$), with a Gaussian distribution with $\mu_3 < 1$, well below the LGI violation condition. For $K_4$ the corresponding condition, $K_4 > 2$, is never fulfilled.

\subsection{Pseudodata}\label{subsec_pseudodata}
Looking for a clearer indication of the LGI violation in the experiments considered here, we implemented a common procedure to increase the statistics and to consider additional information from the experimental data and the BF predictions, i.e., the uncertainties on the measured energy and the oscillation probability.

In this case, we generate a set of pseudodata normally distributed around the measurement ($\psi_i, P_{\nu_{\alpha}\to\nu_{\alpha},i}^{\rm{exp}}$) and the BF prediction, ($\psi_i, P_{\nu_{\alpha}\to\nu_{\alpha},i}^{\rm{BF}}$) to compute $K_{3,4}$. For the experimental observations, the normal distributions are centered in the data points (red dots in Figs.~\ref{fig_Prob_React} and \ref{fig_Prob_Accel}), with horizontal and vertical widths given by the bin width and the experimental error, respectively. For the BF prediction, we use the same location of the phase (or $L/E$) and widths, and the probability is centered at the height of the black line.

\begin{figure}
 \centering
 \includegraphics[width=0.9\linewidth]{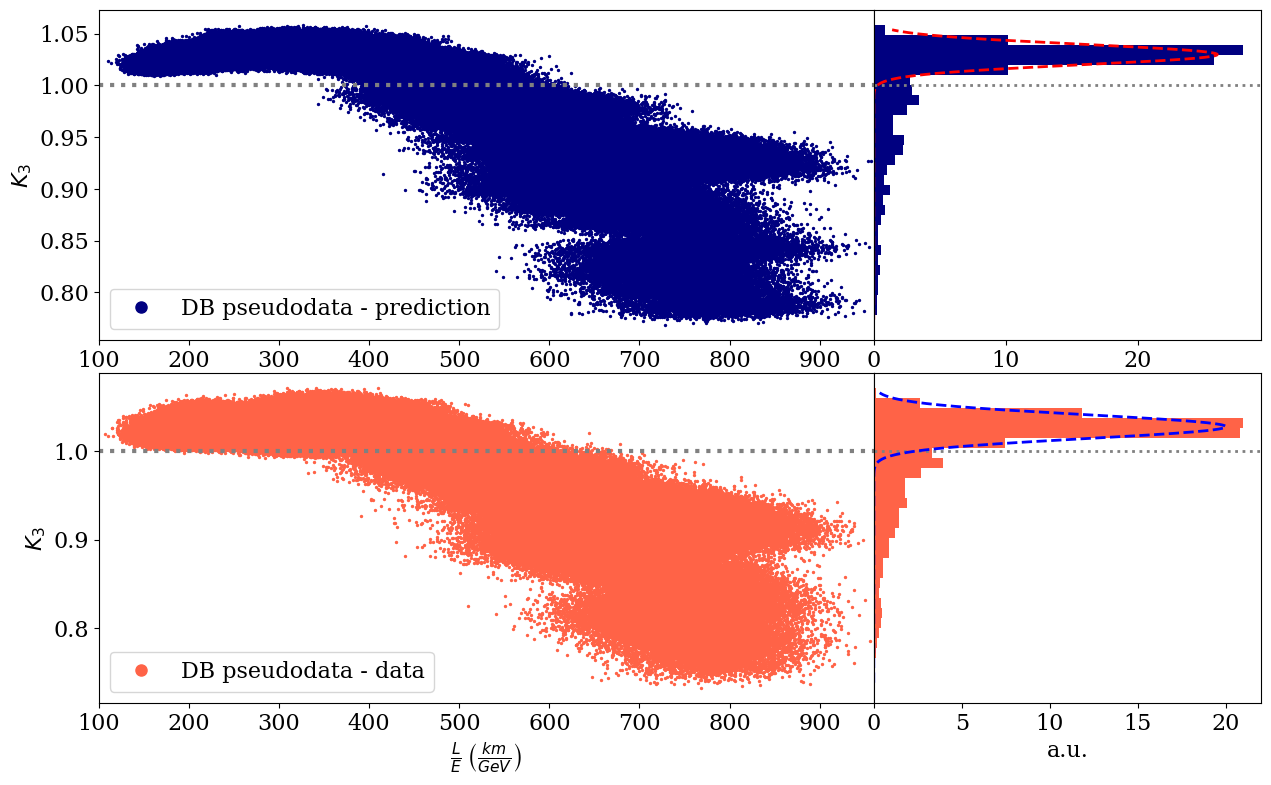}
 \includegraphics[width=0.9\linewidth]{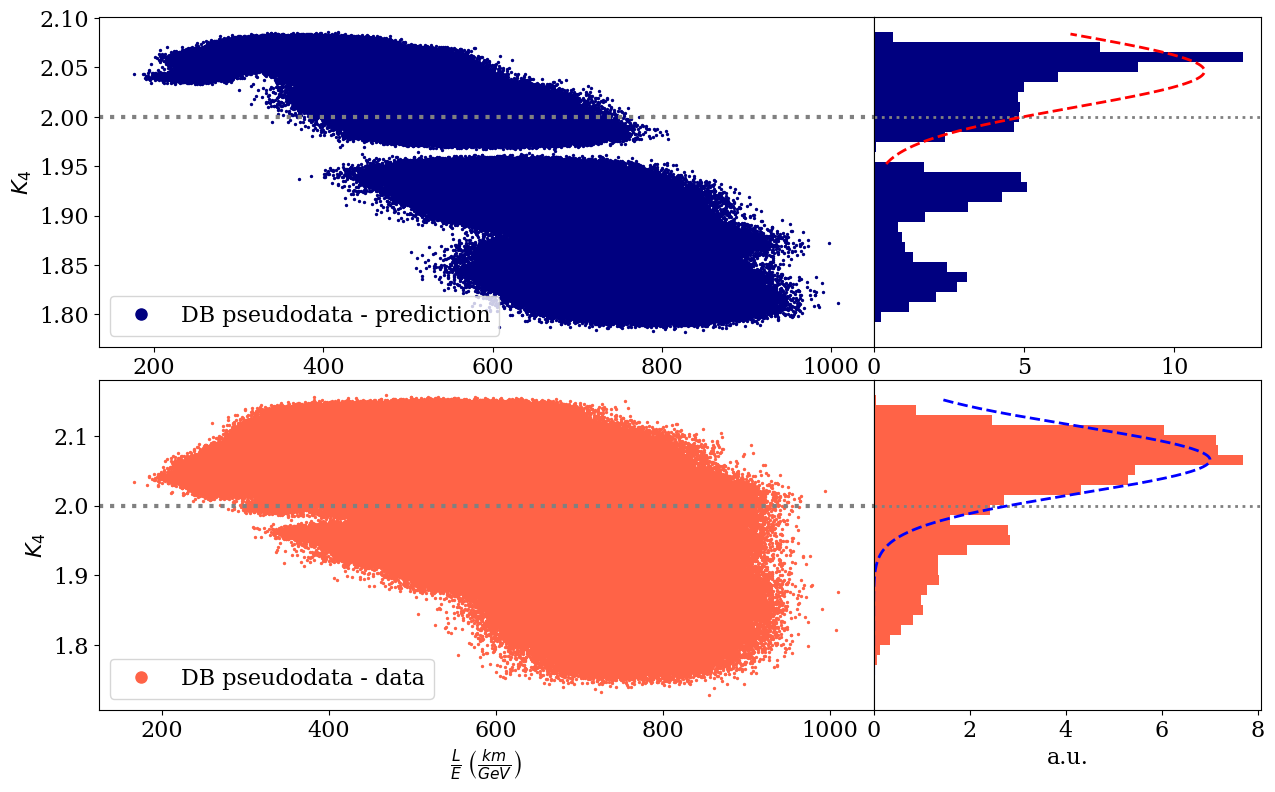}
 \caption{Generated pseudodata satisfying the LGI condition for $K_3$ obtained from the Daya Bay prediction (top panels) and the observations (bottom panels). The corresponding $K_3$ distributions are plotted as histograms on the right hand side, together with the fitted Gaussian.}
 \label{fig_k3BD_pseudo}
\end{figure}

\begin{figure}
 \centering
 \includegraphics[width=0.9\linewidth]{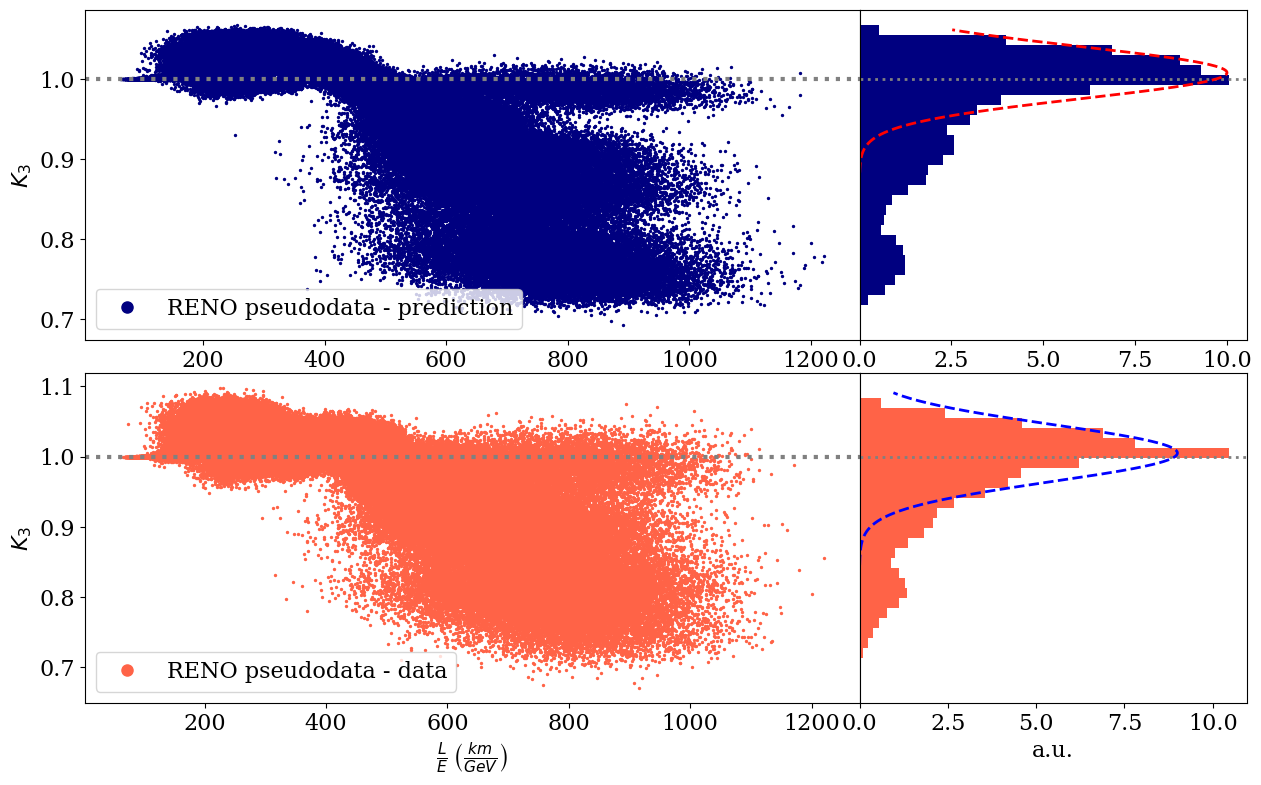}
 \includegraphics[width=0.9\linewidth]{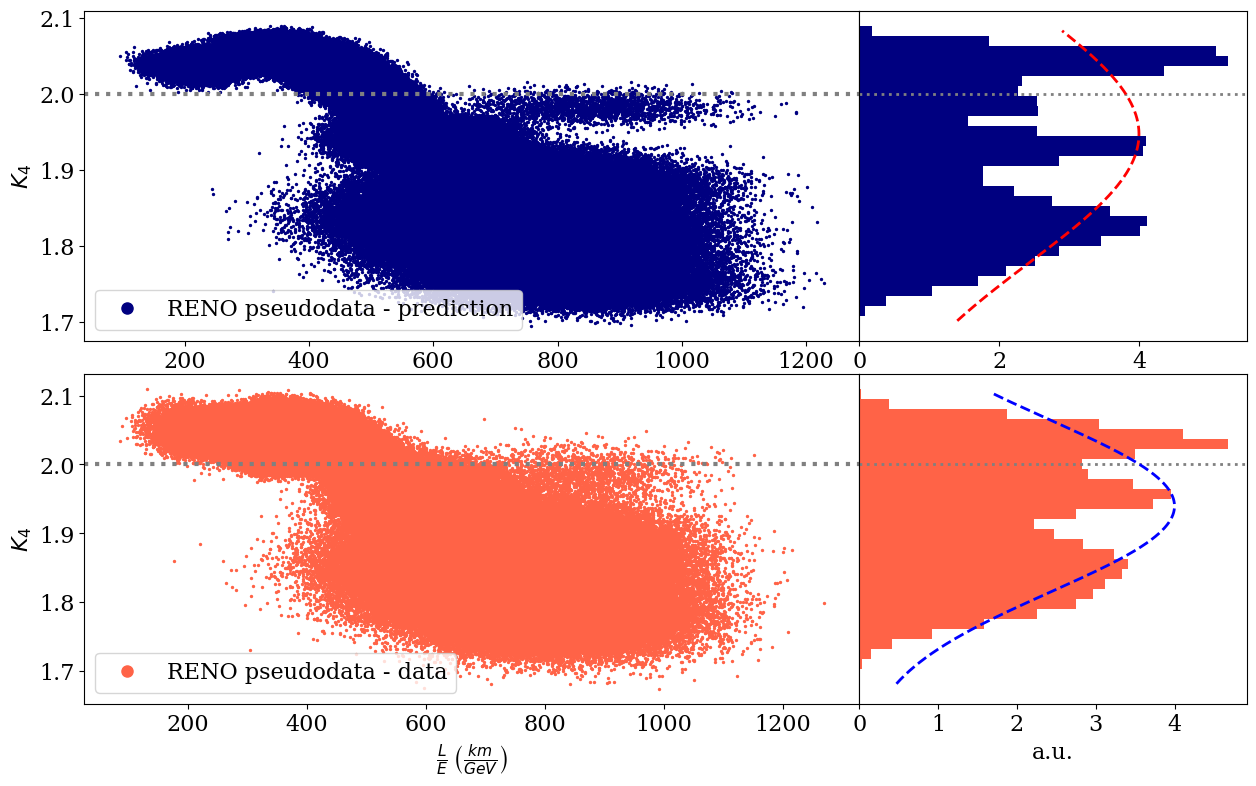}
 \caption{As in Fig.~\ref{fig_k3BD_pseudo}, but for the RENO experiment.}
 \label{fig_k3RENO_pseudo}
\end{figure}

\begin{figure}
 \centering
 \includegraphics[width=0.9\linewidth]{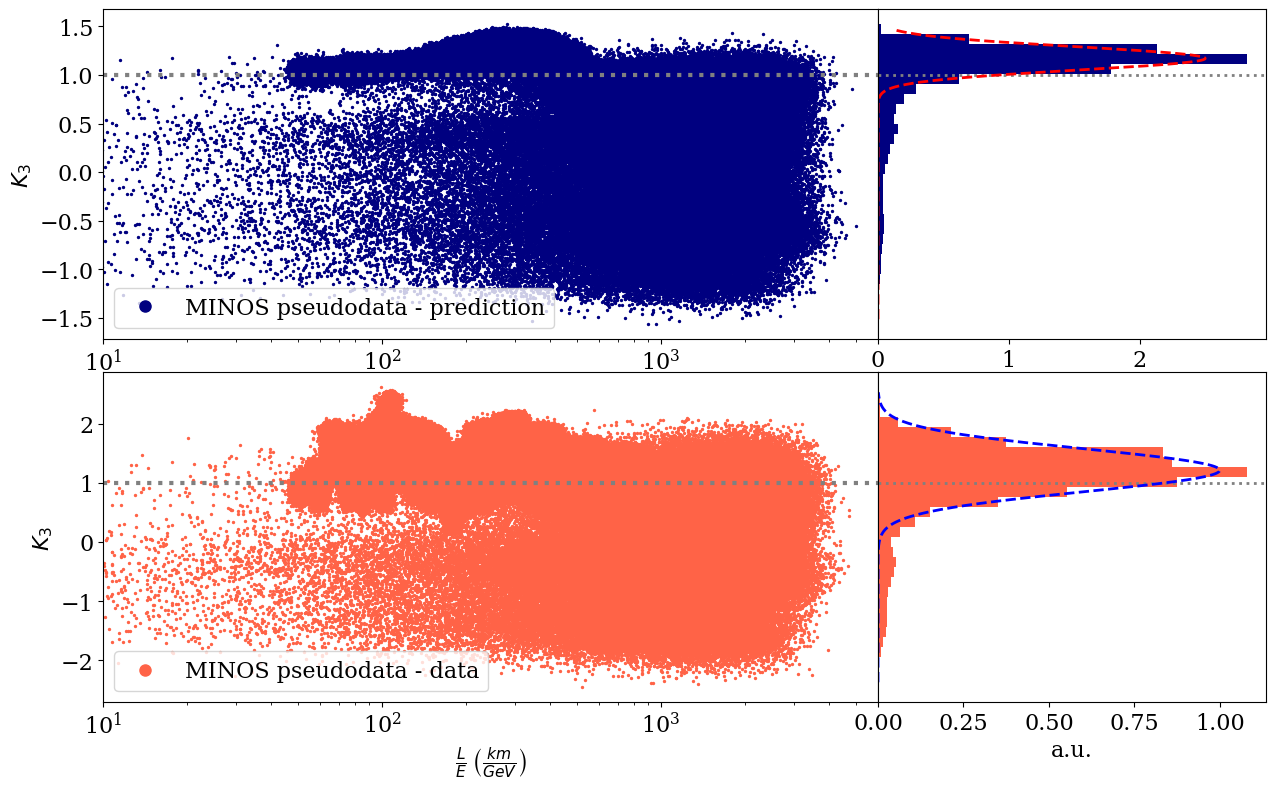}
 \includegraphics[width=0.9\linewidth]{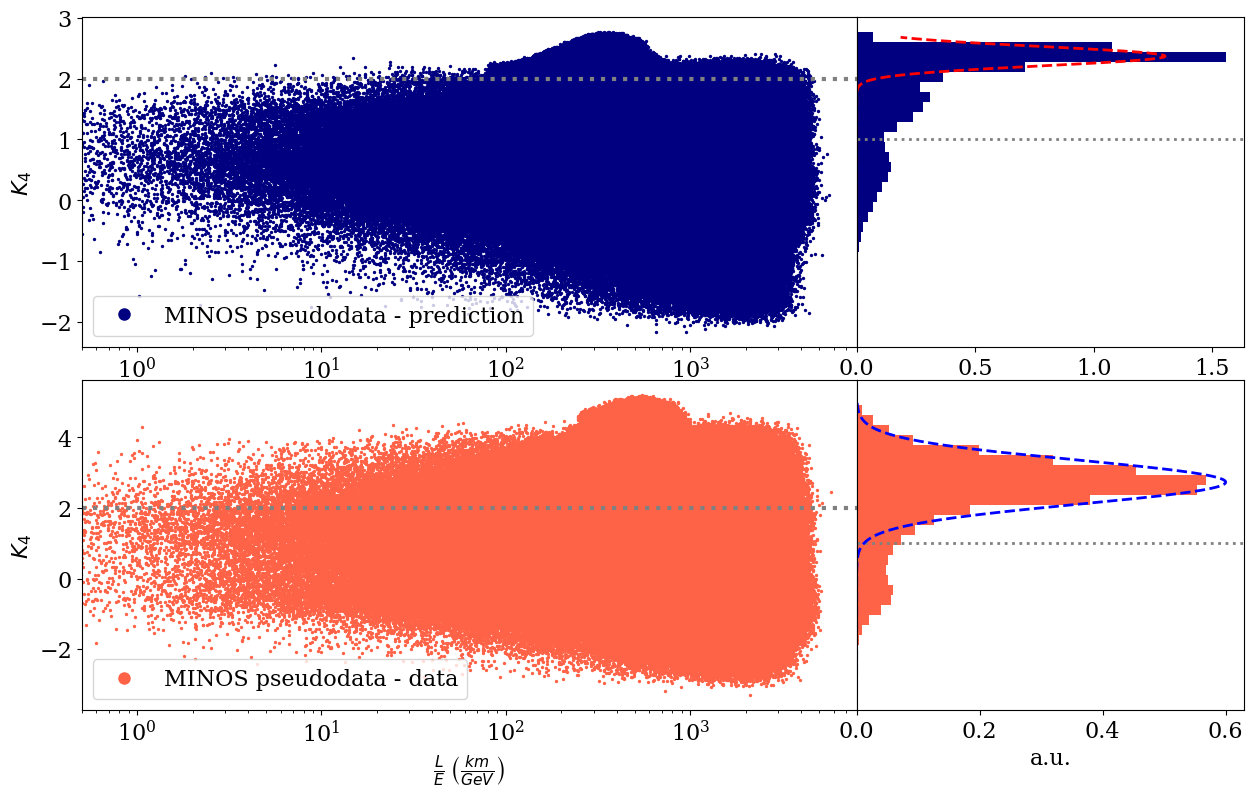}
 \caption{As in Fig.~\ref{fig_k3BD_pseudo}, but for the MINOS experiment.}
 \label{fig_k3MINOS_pseudo}
\end{figure}

\begin{figure}
 \centering
 \includegraphics[width=0.9\linewidth]{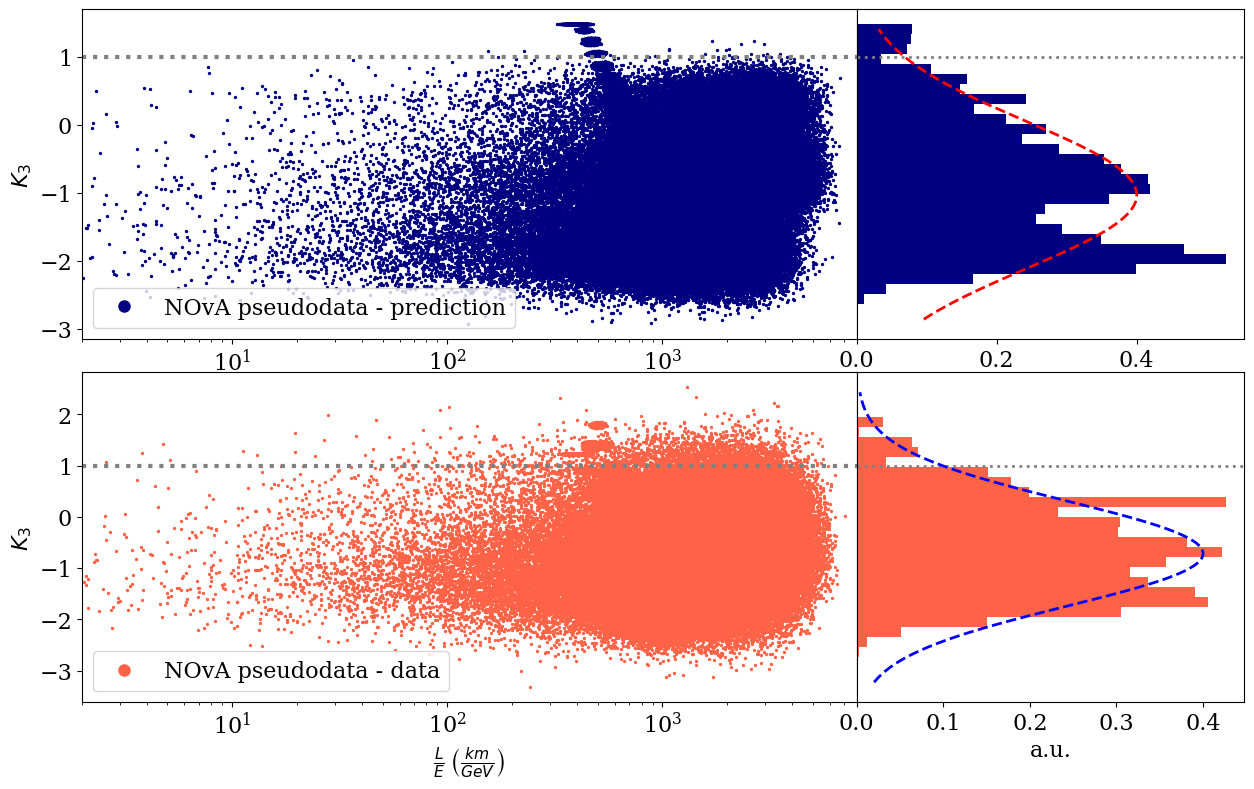}
 \includegraphics[width=0.9\linewidth]{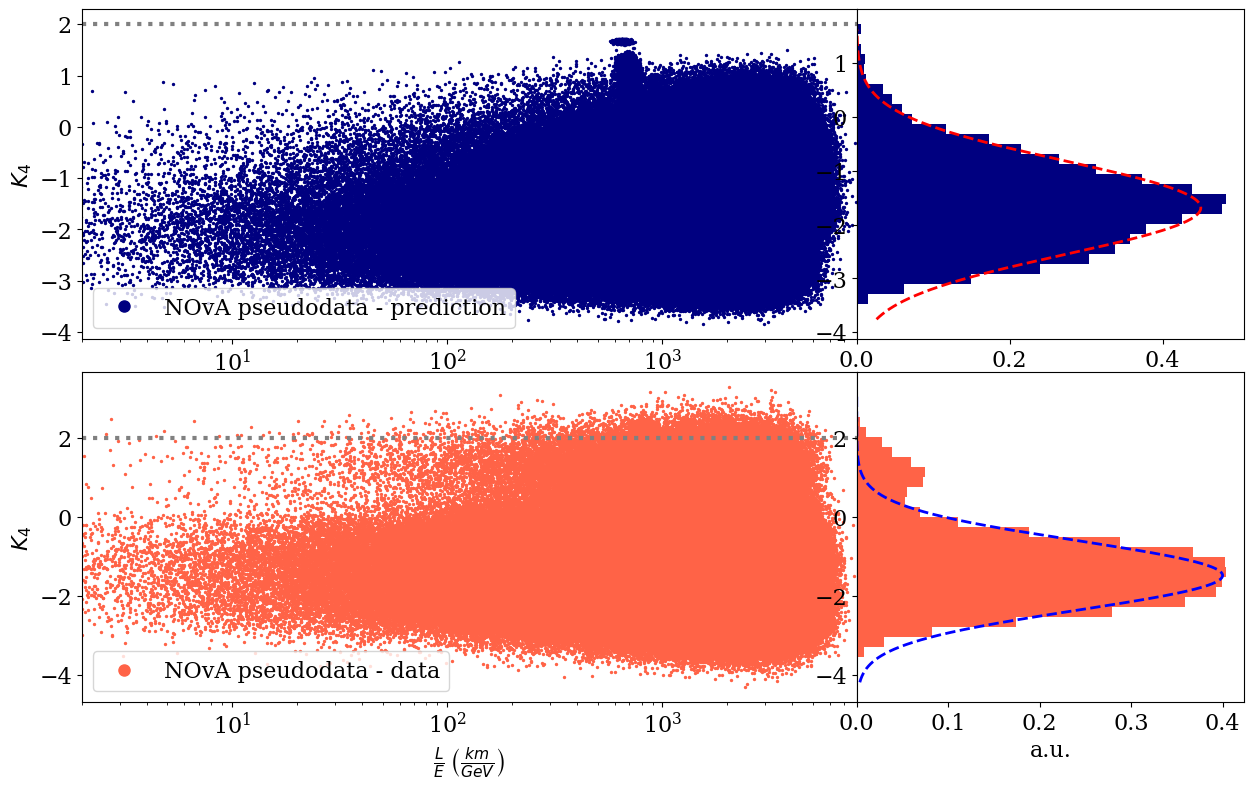}
 \caption{As in Fig.~\ref{fig_k3BD_pseudo}, but for the NOvA experiment.}
 \label{fig_k3NOvA_pseudo}
\end{figure}

Following this description, we generate 2000 pseudodata at each phase, compute the value of $K_{3,4}$ and evaluate the corresponding LGI violation condition. The distribution of the accepted points (those that follow the sum rules mentioned in Sec.~\ref{sec_LGI}) are shown in Figs.~\ref{fig_k3BD_pseudo}--\ref{fig_k3NOvA_pseudo}, while the resulting number of LGI violations and the mean and width of a Gaussian fit to the distributions are presented in Table \ref{tab:LGIv_pseudo}. Although, thanks to the large number of points, fitting a Gaussian distribution to all histograms was possible (hence the means and standard deviations reported in Table \ref{tab:LGIv_pseudo}), we see that some of the distributions are not normal and the corresponding standard deviations (see Table \ref{tab:LGIv_pseudo}) reflect this.


\begin{table}
\caption{Number of LGI violations obtained from analysis of the generated pseudodata for each experiment.} \label{tab:LGIv_pseudo}
\begin{tabular}{@{}llcccccc@{}}
\br
Experiment & & $K_3 > 1\,(\%)$ & $\mu_3$ & $\sigma_3$ & $K_4 > 2\,(\%)$ & $\mu_4$ & $\sigma_4$ \\
\mr
\multirow{2}{*}{DB}    & BF Pred. & 75.2 & 1.03 & 0.01 & 53.1 & 2.05 & 0.04 \\
                       & Data     & 73.2 & 1.03 & 0.01 & 70.6 & 2.07 & 0.05 \\\mr
\multirow{2}{*}{RENO}  & BF Pred. & 44.5 & 1.01 & 0.03 & 27.5 & 1.95 & 0.17 \\
                       & Data     & 44.2 & 1.00 & 0.04 & 27.9 & 1.94 & 0.13 \\\mr
\multirow{2}{*}{MINOS} & BF Pred. & 77.4 & 1.17 & 0.12 & 59.9 & 2.37 & 0.15 \\
                       & Data     & 67.2 & 1.24 & 0.36 & 76.8 & 2.73 & 0.62 \\\mr
\multirow{2}{*}{NOvA}  & BF Pred. & 3.5 & -1.04 & 1.10 & 0.0 & -1.69 & 0.87 \\
                       & Data     & 3.6 & -0.72 & 1.03 & 0.4 & -1.48 & 0.88 \\
\br
\end{tabular}
\end{table}
Again, the LGI violation is apparent for all experiments except NOvA. It is clear from Figs.~\ref{fig_k3BD_pseudo}-\ref{fig_k3MINOS_pseudo}, as well as from the numbers in Table \ref{tab:LGIv_pseudo}, that a large number of LGI violations are obtained from the generated pseudodata for the BF predictions and the observations. 

DB and MINOS exhibit the highest number of violations: for DB, more than 73\,\% of the points satisfy $K_3 > 1$ and more than 53\,\% satisfy $K_4 > 2$; in turn, for MINOS more than 60\,\% of the points satisfy $K_3 > 1$ or $K_4 > 2$. It is worth stressing that, although the $K_4$ histograms for DB showed multiple peaks, it was possible to fit a Gaussian distribution with a mean located above $K_4 = 2$, indicating that the most of the generated points are concentrated above this limit. For $K_3$, on the other hand, the corresponding histograms for the points generated at the BF predictions and the observations (upper plots in Fig.~\ref{fig_k3BD_pseudo}) exhibit a single peak distribution, making the Gaussian fit more appealing, with a mean just above $K_3 = 1$, and a width consistent with the fact that more than 70\,\% of the generated points show LGI violation.

For MINOS, the $K_{3,4}$ histograms (Fig.~\ref{fig_k3MINOS_pseudo}) are well fitted with a Gaussian distribution with means well above the LGI violation limits, $K_3 = 1$ and $K_4 = 2$, and widths that reinforce the LGI violation in the vast majority of the generated points.

The information collected in Table \ref{tab:LGIv_pseudo} shows that the number of LGI violations in the case of RENO is $\lesssim 45\,\%$ for $K_3$, and $\sim 30\,\%$ for $K_4$. The corresponding distributions (right panels of Fig.~\ref{fig_k3RENO_pseudo}) have a structure similar to those for DB, with an acceptable Gaussian fit for $K_3$, but not much for $K_4$, with the difference that the LGI violation is less evident in this experiment. The fitted Gaussians have a mean around the LGI violation limit for $K_3$ ($K_3 = 1$) and just bellow $K_4 = 2$, respectively, and the corresponding widths support the LGI violation counting.

Finally, as observed from the analysis of the data and BF predictions in Sec.~\ref{subsec_data}, the study carried out with the NOvA data results in the smallest number of LGI violations (Fig.~\ref{fig_k3NOvA_pseudo} and last two rows of Table \ref{tab:LGIv_pseudo}), up to the point that none of the generated (pseudodata) points violate the LGI when $K_4$ is considered. In addition, in this case it is possible to fit a Gaussian distribution, and the four fits led to means well below the corresponding limits, $K_{3,4} = 1,2$. 

At first glance, these later results appear odd, notably if compared with MINOS (with a clear evidence of LGI violation), as both are LBL accelerator experiments. However, the particular differences of the experimental setup as well as their results, might provide an insight about this supposed tension. Regarding the experimental setups, as mentioned in Sec.~\ref{sec_Experiments}, the baseline and energy range are different, the latter being the most important aspect here, since it is the larger energy range explored by MINOS the factor that allows us to explore many more phases. This is evident by looking at the left panel of Fig.~\ref{fig_Prob_Accel}, where the phases expand in the interval $\frac{L}{E} \sim (10,2000)\, \frac{\rm{km}}{\rm{GeV}}$, and comparing them with those of NOvA, right panel of Fig.~\ref{fig_Prob_Accel}, where the phase interval is shorter, $\frac{L}{E} \sim (100,2500)\, \frac{\rm{km}}{\rm{GeV}}$.

\begin{figure}[ht!]
 \centering
 \includegraphics[width=\linewidth]{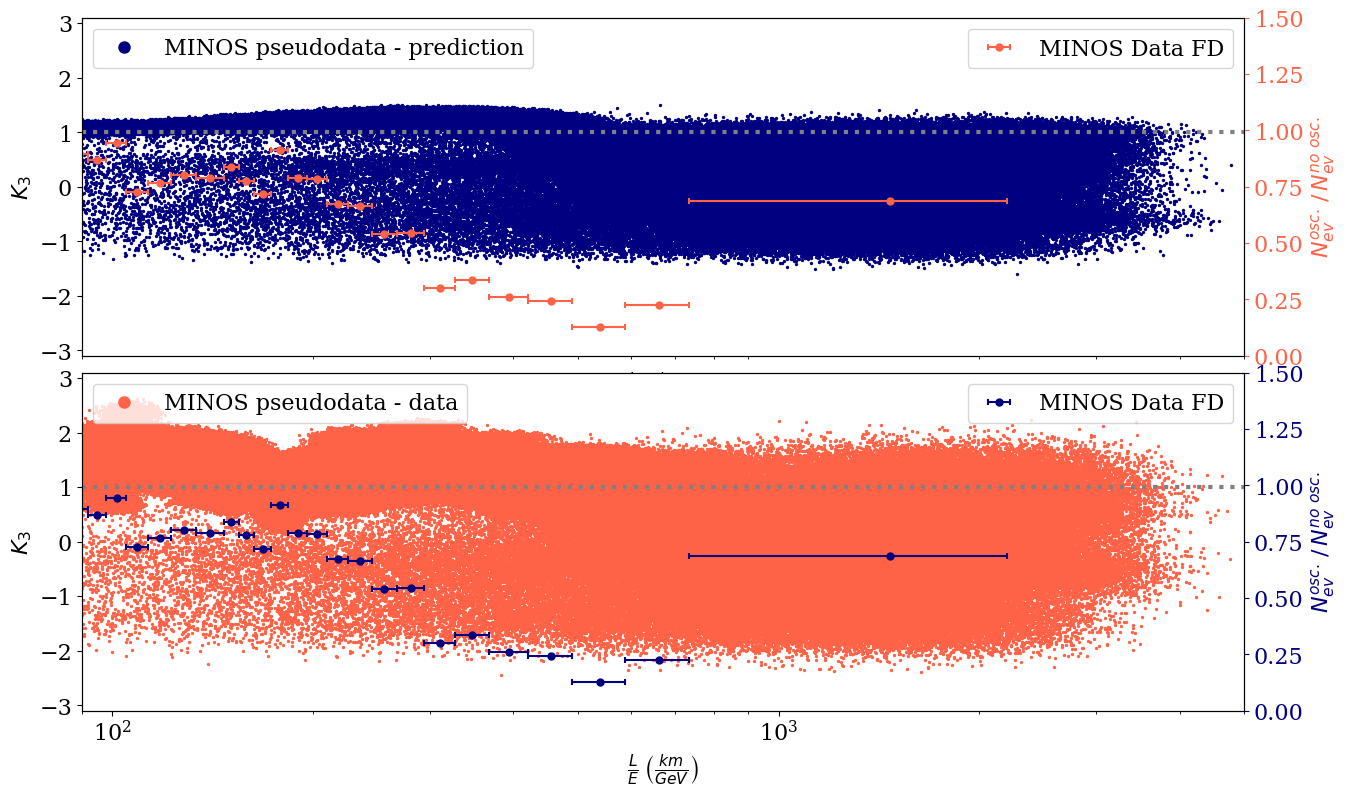}
 \includegraphics[width=\linewidth]{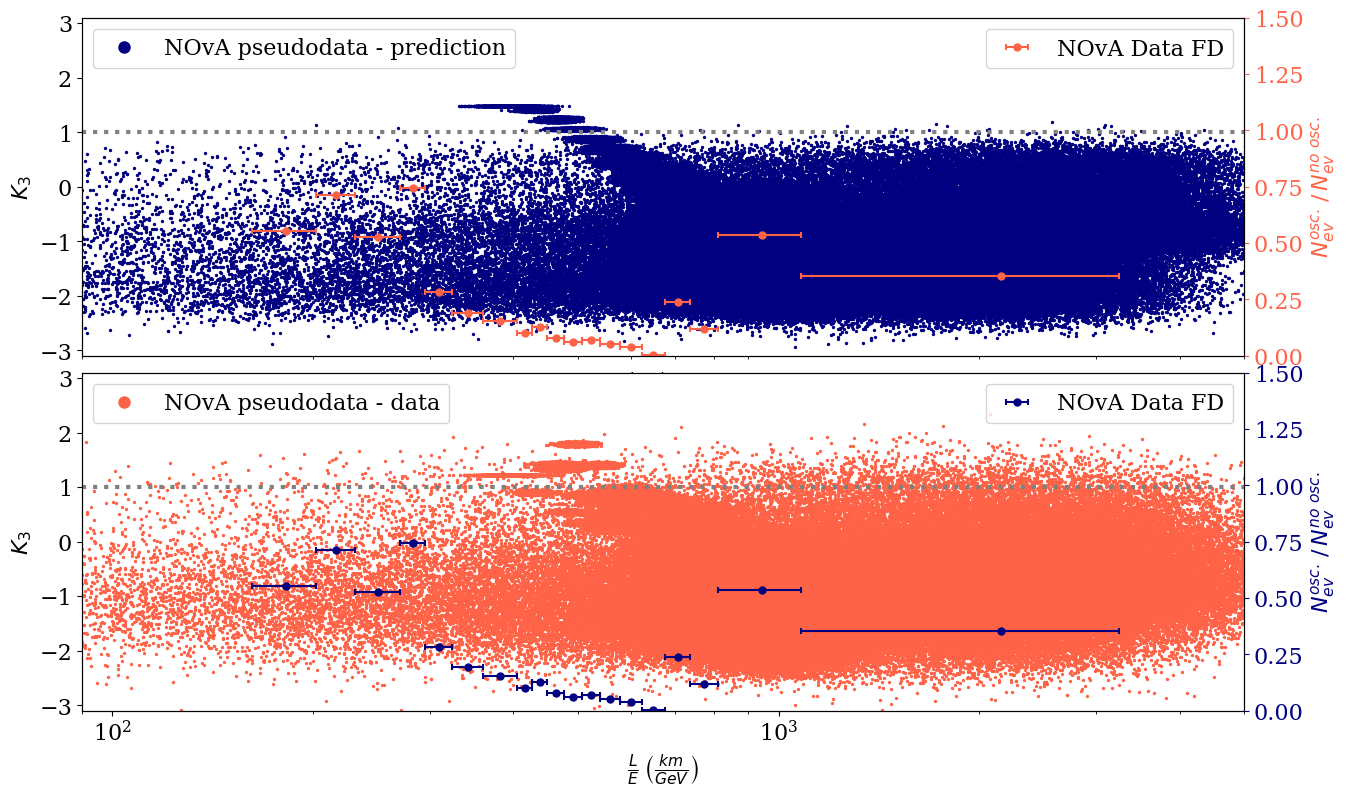}
 \caption{Comparison of the points satisfying the LGI sum rule for $K_3$ (left vertical axis) between MINOS (top) and NOvA (bottom), including the measured event ratio (right vertical axis).}
 \label{fig_k3_zoom}
\end{figure}

To make the comparison more clear, we display again $K_3 \, \rm{vs.} \, L/E$ for the two accelerator experiments in Fig.~\ref{fig_k3_zoom}, setting the same range in the horizontal axis. This time we also include the measured ratio of observed over expected events (without errors; right vertical axis). Given that MINOS also reports measurements for $L/E < 100 \, \frac{\rm{km}}{\rm{GeV}}$ (not shown in Fig.~\ref{fig_k3_zoom}), the high density population of generated pseudodata expands throughout the entire displayed interval, with the corresponding value of $K_3$ (left vertical axis); meanwhile, all NOvA experimental data are visible in these plots, so when the pseudodata are generated, the density of points at $L/E \lesssim 300 \, \frac{\rm{km}}{\rm{GeV}}$ is much lower, and the vast majority lay below $K_3 = 1$. In fact, this same circumstance (i.e., NOvA having fewer data points than MINOS) also explains why there are fewer sets of phases, $\psi_i, \, i=1,2,3$, that fulfill the required sum rule, hence the lower number of pseudodata for NOvA for the whole $L/E$ range displayed in the figure.

The other important feature that one can see from Fig.~\ref{fig_k3_zoom} (as well as from Fig.~\ref{fig_Prob_Accel}), is that the ratio of events (proportional to the oscillation probability) measured by NOvA is systematically lower than that measured by MINOS. This implies that, when $K_3$ is computed by means of Eq.~\ref{eq:k3_python}, the results are noticeable lower for NOvA than for MINOS. What emerges from this is that, in the region where both experiments have a large number of pseudodata generated ($\frac{L}{E} \sim (300,2000)\, \frac{\rm{km}}{\rm{GeV}}$), most of the points for NOvA lay below the LGI violation limit, $K_3 = 1$. Then one can conclude that, although the NOvA results appear to indicate that the LGI violation is weak, the outcome is really consistent with that of MINOS. In addition, it is in this same region that $K_3$ takes the lowest values in the MINOS analysis (and $K_3 < 1$). Actually, the compatibility goes beyond: looking at Figs.~\ref{fig_k3BD_pseudo} and \ref{fig_k3RENO_pseudo} for DB and RENO, respectively, one can notice that $K_3 < 1$ (no LGI violation) is obtained for $\frac{L}{E} \sim (300,1000)\, \frac{\rm{m}}{\rm{MeV}}$, which is essentially the same interval for NOvA and MINOS (scaled by a factor $10^3$, both in $L$ and $E$).

Therefore, in summary, Daya Bay and MINOS exhibit a strong indication of LGI violation, while for RENO and (particularly for) NOvA, the manifestation is weaker. What we have observed and explained above is that the data collected by these two experiments are concentrated in a region where $\psi_i$ and their corresponding oscillation probability, lead to a small value of the LGI parameter ($K_{3,4} < 1,2$). In addition, all the analyzed experiments show that the larger the phase ($\psi_i \sim (L/E)_i$), the lower $K_{3,4}$, which coincides with the results presented in Ref.~\cite{Wang:2022tnr}, in particular with regard to the analysis of KamLAND data presented in that reference. 

\begin{figure}[ht!]
 \centering
 \includegraphics[width=0.8\linewidth]{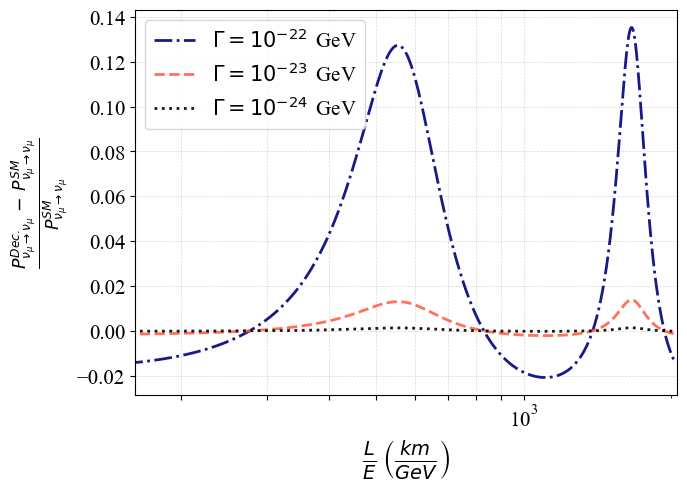}
 \caption{Comparison of the $\nu_{\mu}$ survival probability given by a model including decoherence effects (parameterized by the factor $\Gamma$) with the standard prediction. Three different values of the decoherence parameter $\Gamma$ are used, showing that with a small enough $\Gamma$, the two predictions become indistinguishable. These curves are generated using the model described in \cite{Oliveira:2013nua} as implemented in the \texttt{OscProb} package \cite{Coelho:OscProb}.}
 \label{fig_OscProb_Decoh}
\end{figure}

As long as the LGI is implemented as a test of quantum mechanics in macroscopic systems and phenomena (e.g.~neutrino oscillations), our results suggest that it is necessary to consider data with a low $L/E$ ratio. 

In terms of testing the coherence of neutrino oscillations by means of the LGI, our results for NOvA data could be interpreted as a possible indication of decoherence effects. In fact, considering a (simplified\footnote{We only consider one of the decoherence parameters to be different from zero.} version of the) model of three neutrinos with dissipation effects as studied in \cite{Oliveira:2013nua}, the $\nu_{\mu}$ survival probability is found to be comparable to the standard prediction (with coherent mass eigenstates evolution), as long as the decoherence parameter is small enough. This is shown in Figure \ref{fig_OscProb_Decoh}, where the standard survival probability, $P^{\rm{SM}}$, is compared with the probability including decoherence effects, for three different values of the corresponding parameter $\Gamma$. If $\Gamma$ is large ($\gtrsim \mathcal{O}(10^{-22})$ GeV, blue dash-dotted line), the decoherence model differs considerably from the standard one; yet, for $\Gamma \lesssim 10^{-23}$ GeV, the two predictions become almost indistinguishable, indicating that NOvA data could be well reproduced by a model that includes decoherence effects.


\section{Conclusions}\label{sec_Conclusion}
In this work we analyzed the violation of the Leggett-Garg inequality in the context of neutrino oscillation experimental results by the Daya Bay and RENO reactor experiments, and the MINOS and NOvA LBL accelerator experiments. A strong manifestation of LGI violation is observed in the DB and MINOS data, while for the RENO and NOvA data the indication is weaker, especially for NOvA.

Having a fixed baseline, it is apparent that the differences in the results obtained from MINOS and NOvA come from the particular energy range (and hence the number of data points) of each one: since MINOS can explore neutrino energies up to $\sim$50 GeV, it is possible to probe smaller phases in the search for LGI violations than with NOvA data (with $E \lesssim 5$ GeV). On the other hand, while DB and RENO data were collected using similar ranges of distances and energies, it is the data from the smaller $L/E$ region (i.e. EH1 and EH2 for DB and ND for RENO, where the former has many more data points than the later --see Fig.~\ref{fig_Prob_React}--) that shows the strongest signal of LGI violations. 

Our results then demonstrate that the evidence for LGI violation is weaker for larger phases $\psi \sim L/E$. This is observed in all the data sets considered in this study, an aspect that shows the consistency of this analysis and that provides a compelling and relevant component to be considered when searching for evidences of quantum mechanical decoherence on neutrino oscillations.

\section*{Acknowledgments}
We are grateful to our friends and colleagues who attended the \emph{Neutrinos en Colombia 2023} Workshop\footnote{\url{https://indico.cern.ch/event/nuco2023}} and the 8th Colombian Meeting on High Energy Physics (COMHEP 2023\footnote{\url{https://indico.cern.ch/event/comhep8}}), where preliminary results of this work were presented, for their relevant comments and great support. We thank also David V.~Forero for the insightful discussion during the preparation of this manuscript. M.A.A.~thanks Alexis A. Aguilar-Arevalo for his thorough review of the manuscript, as well as for his sharp and deep comments and suggestions that helped us improve the presented discussion.

\section*{Data availability statement}
Availability of data and materials. Data used in our work are freely accessible and published by the corresponding collaborations. We give proper credit and include the necessary references to these works; hence no further data deposit is needed.


\end{document}